\RequirePackage[OT1]{fontenc} 
\documentclass[conference]{IEEEtran}
\IEEEoverridecommandlockouts
\usepackage[]{authblk}

\usepackage{graphicx,import}
\graphicspath{{img/}}
\usepackage{url}
\usepackage[numbers,sort&compress]{natbib}
\usepackage{booktabs}
\usepackage{array}
\usepackage{multirow}

\usepackage[inline]{enumitem}

\usepackage{pgfplots}
\usepgfplotslibrary{groupplots,dateplot}
\usetikzlibrary{patterns,shapes.arrows}
\pgfplotsset{compat=newest}
\usetikzlibrary{spy}
\usepgfplotslibrary{polar}
\usetikzlibrary{positioning}
\pgfplotsset{every axis plot/.append style={thick}}

\usepackage{tikz}
\usetikzlibrary{external}
\usepackage{float} 
\usepackage{placeins}

\usepackage{subcaption}
\usepackage{amsmath,amssymb,amsfonts,bm}
\usepackage{graphicx}
\usepackage{url}
\usepackage{xcolor}
\usepackage{tabularx}
\usepackage[binary-units=true]{siunitx}
  \DeclareSIUnit{\belmilliwatt}{Bm}
\DeclareSIUnit{\dBm}{\deci\belmilliwatt}

\usepackage{listings}
\usepackage{comment}
\usepackage{import}

\usepackage[referable]{threeparttablex}
\usepackage{enumitem}
\usepackage[referable]{threeparttablex}
\renewlist{tablenotes}{enumerate}{1}
\makeatletter
\setlist[tablenotes]{label=\tnote{\alph*},ref=\alph*,itemsep=\z@,topsep=\z@skip,partopsep=\z@skip,parsep=\z@,itemindent=\z@,labelindent=\tabcolsep,labelsep=.2em,leftmargin=*,align=left,before={\footnotesize}}
\makeatother

\newcommand*{\vectr}[1]{\MakeLowercase{\bm{\mathit{#1}}}}

\newcommand{\norm}[1]{\left\lVert#1\right\rVert}
\newcommand{\abs}[1]{\left\lvert#1\right\rvert}
\newcommand{\expt}[1]{\mathbb{E} \left\{#1\right\}}
\newcommand{\var}[1]{\mathbb{V} \left\{#1\right\}}
\newcommand{\normalized}[1]{\bar{#1}}

\newcommand{\fredrik}[1]{{\color{orange}[Fredrik]}}

\definecolor{colorULA}{rgb}{0.12157,0.46667,0.70588}
\definecolor{colorURA}{rgb}{1.00000,0.49804,0.05490}
\definecolor{coloriid}{rgb}{0.17255,0.62745,0.17255}
\definecolor{colorIID}{rgb}{0.17255,0.62745,0.17255}
    
\begin{document}

\title{Massive MIMO goes Sub-GHz: Implementation and Experimental Exploration for LPWANs}

\author[*]{Gilles Callebaut}
\author[* $\dag$]{Sara Gunnarsson}
\author[*]{Andrea P. Guevara}
\author[$\dag$]{\\Fredrik Tufvesson}
\author[*]{Sofie Pollin}
\author[* $\dag$]{Liesbet Van der Perre}
\author[$\dag$]{Anders J Johansson \thanks{This work was funded by the European Union's Horizon 2020 under grant agreement no.~732174 (ORCA project) and no.~731884 (IoF2020 program - IoTrailer use case).\\
This work has been accepted by IEEE Asilomar 2020.}}
\affil[*]{Department of Electrical Engineering, KU Leuven, Belgium}
\affil[$\dag$]{Department of Electrical and Information Technology, Lund University, Sweden}

\maketitle

\begin{abstract}
Low-Power Wide-Area Networks operating in the unlicensed bands are being deployed to connect a rapidly growing number of Internet-of-Things devices. While the unlicensed sub-GHz band offers favorable propagation for long-range connections, measurements show that the energy consumption of the nodes is still mostly dominated by the wireless transmission affecting their autonomy. We investigate the potential benefits of deploying massive MIMO technology to increase system reliability and at the same time support low-energy  devices with good coverage at sub-GHz frequencies. The impact of different antenna configurations and propagation conditions is analyzed. Both actual average experienced array gain and channel hardening are examined. 
The assessment demonstrates the effect of channel hardening as well as the potential benefits of the experienced array gain. These measurements serve as a first assessment of the channel conditions of massive MIMO at sub-GHz frequencies and are, to the best of our knowledge, the first of its kind.
\end{abstract}

\begin{IEEEkeywords}
Internet-of-Things, LPWAN, Massive MIMO, Sub-GHz, Testbed
\end{IEEEkeywords}\FloatBarrier

\section{Introduction}\label{sec:introduction}
The Internet-of-Things (IoT) is strongly  
growing. A variety of applications -- each having its own requirements -- are being deployed in a fast pace. Consequently, the number of IoT devices are increasing exponentially. Furthermore, a high number of all the applications demand that these devices are autonomous for a couple of years. Low-Power Wide-Area Networks (LPWANs) operating in the unlicensed sub-GHz spectrum are being deployed to connect these devices. However, the current LPWANs are unable to meet the new stringent requirements of a new wave of IoT demanding low-power, massive connectivity and high reliability. 

Massive MIMO is a technology where a large number of antennas is used at the base station to serve multiple devices at the same time. By introducing more antennas, the reliability, coverage, energy-efficiency and the number of connected nodes can be enhanced in comparison to when using conventional single-antenna gateways. Hence, there is a potential benefit of using massive MIMO in the unlicensed sub-GHz spectrum to support the growth of IoT applications.

In order to validate the anticipated benefits of using massive MIMO for LPWANs, testbeds or other measurement equipment is required. In Table~\ref{tab:overview-testbeds} an overview of current available massive MIMO testbeds operating at sub-6GHz is presented. In contrast to LPWAN applications, most of these testbeds are designed for high throughput. The trend is also to rather explore higher frequencies and therefore a lot of attention is given to designing equipment to operate in the mmWave, or even terahertz, bands. For instance, Facebook introduced Terragraph~\cite{choubey2016introducing}, which is operating at \SI{60}{\giga\hertz}. Although some of the testbeds can operate at sub-GHz frequencies, massive MIMO measurements in this unlicensed band are still lacking and the actual benefit of deploying massive MIMO to support future LPWANs have until now remained unexplored.


To cover this gap, we extended a massive MIMO testbed to be narrow-band, operate in the unlicensed sub-GHz band and designed a versatile antenna array.
Secondly, we conducted the first sub-GHz massive MIMO measurement campaign to collect channels in an outdoor scenario with two different array configurations and made them available open-source\footnote{\url{dramco.be/massive-mimo/measurement-selector/#Sub-GHz}\label{data}}. 
An initial assessment of the radio propagation characteristics were studied, based on the average array gain and channel hardening, showcasing the potential gain of using massive MIMO to evolve current LPWANs in order to pave the way for the new wave of IoT.





The structure of this paper is as follows. In Section~\ref{sec:array} the versatile antenna array is described, followed by a section elaborating on the implementations made in the framework and in the setup in order to comply to the regulations in the unlicensed band. Section~\ref{sec:scenario} presents the measurement scenario and Section~\ref{sec:results} shows the results from the experimental exploration. Finally, the conclusions are given.

\begin{table*}[tbp]

    \centering
    \caption{Overview of massive MIMO testbeds grouped by frequency band capability.}%
    \label{tab:overview-testbeds}
    \begin{tabularx}{\textwidth}{lXXX}
    \toprule
    University/Company & Location & Number of antennas & Frequency range\\ \midrule
    
    \textbf{Sub-GHz} \\
    \hspace{1ex} Eurecom (OAI)~\cite{7955940} & Sophia Antipolis, France& 64 & depends on RF platform\\
    
    \hspace{1ex} KU Leuven~\cite{martinez2018experimental} & Leuven, Belgium  & 64 & \SI{400}{\mega\hertz} to \SI{4.4}{\giga\hertz} \\


      
    \hspace{1ex} Rice University (Argos)~\cite{shepard2017argosv3} & Houston, Texas, USA & 64 (v1) 96 (v2) flexible (v3) & \SI{50}{\mega\hertz} to \SI{3.8}{\giga\hertz} (v3)\\
    
    \textbf{Sub-6GHz} \\
    \hspace{1ex} Norwegian University of Science
and Technology~\cite{testbednorway} & Trondheim, Norway & 64 & \SI{1.2}{\giga\hertz} to \SI{6}{\giga\hertz}\\

\hspace{1ex} University of Lund (LuMaMi)~\cite{kulmami_journal}  & Lund, Sweden &  100 & \SI{1.2}{\giga\hertz} to \SI{6}{\giga\hertz} \\

    
    \hspace{1ex} University of Bristol~\cite{7145598} & Bristol, UK &  128 & \SI{1.2}{\giga\hertz} to \SI{6}{\giga\hertz}\\

    \hspace{1ex} Southeast university~\cite{8246333} & Nanjing, China & 128 & \SI{1.2}{\giga\hertz} to \SI{6}{\giga\hertz}\\
    
    \bottomrule
    \end{tabularx}
\end{table*}


\begin{figure}[tbp]
    \centering
    \input{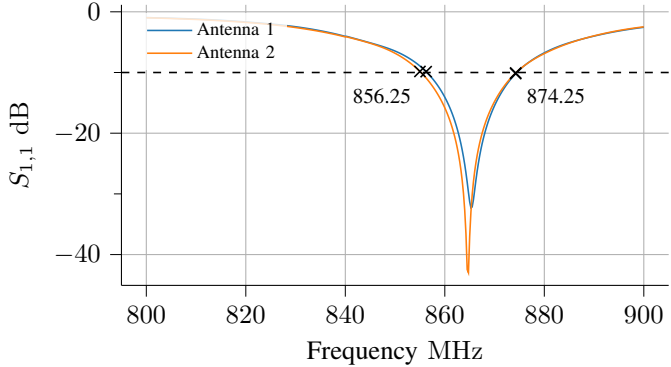}
    \caption{Measured \(S_{1,1}\) parameters of two antennas in the same holder.}%
    \label{fig:s11}
\end{figure}

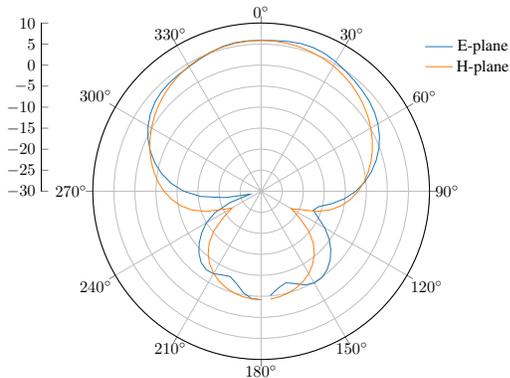
\begin{figure}[tbp]
    \centering
    \resizebox{.8\linewidth}{!}{
\begin{tikzpicture}

\definecolor{color0}{rgb}{0.12157,0.46667,0.70588}
\definecolor{color1}{rgb}{1.00000,0.49804,0.05490}

\begin{polaraxis}[
width = \linewidth,
   xticklabel=$\pgfmathprintnumber{\tick}^\circ$,
  xtick={0,30,...,330},
  ytick={-30,-25,...,10},
    y coord trafo/.code=\pgfmathparse{#1+30},
   rotate=-90,
    y coord inv trafo/.code=\pgfmathparse{#1-30},
   x dir=reverse,
  xticklabel style={anchor=-\tick-90},
  yticklabel style={anchor=east, xshift=-4.75cm},
  y axis line style={yshift=-4.75cm},
  ytick style={yshift=-4.75cm},
  legend style={fill opacity=0.8, draw opacity=1, text opacity=1, at={(0.03000,0.97000)}, anchor=north west, draw=white!80.00000!black},
  legend style={nodes={scale=1, transform shape}},
  legend style={draw=none},
]
\addplot [semithick, color0]
table [row sep=\\] {%
-180.0 -4.197\\
-175.0 -4.345\\
-170.0 -5.519\\
-165.0 -7.291\\
-160.0 -8.402999999999999\\
-155.0 -8.037\\
-150.0 -7.332000000000001\\
-145.0 -7.21\\
-140.0 -7.83\\
-135.0 -9.082\\
-130.0 -10.78\\
-125.00000000000001 -12.74\\
-119.99999999999999 -14.98\\
-115.0 -17.82\\
-110.0 -22.14\\
-105.00000000000001 -27.18\\
-100.0 -21.31\\
-95.0 -15.6\\
-90.0 -11.62\\
-85.0 -8.54\\
-80.0 -6.002000000000001\\
-75.0 -3.8310000000000004\\
-70.0 -1.949\\
-65.0 -0.3265\\
-59.99999999999999 1.04\\
-55.0 2.138\\
-50.0 2.955\\
-45.0 3.4939999999999998\\
-40.0 3.801\\
-35.0 3.989\\
-29.999999999999996 4.227\\
-25.0 4.6339999999999995\\
-20.0 5.1610000000000005\\
-14.999999999999998 5.625\\
-10.0 5.877999999999999\\
-5.0 5.917000000000001\\
0.0 5.888999999999999\\
5.0 5.974\\
10.0 6.192\\
14.999999999999998 6.379\\
20.0 6.353\\
25.0 6.0489999999999995\\
29.999999999999996 5.529\\
35.0 4.938\\
40.0 4.4030000000000005\\
45.0 3.9410000000000003\\
50.0 3.465\\
55.0 2.864\\
59.99999999999999 2.065\\
65.0 1.038\\
70.0 -0.2182\\
75.0 -1.694\\
80.0 -3.386\\
85.0 -5.307\\
90.0 -7.499\\
95.0 -10.03\\
100.0 -12.95\\
105.00000000000001 -15.82\\
110.0 -16.8\\
115.0 -15.0\\
119.99999999999999 -12.51\\
125.00000000000001 -10.28\\
130.0 -8.387\\
135.0 -6.807\\
140.0 -5.596\\
145.0 -4.873\\
150.0 -4.782\\
155.0 -5.422999999999999\\
160.0 -6.638\\
165.0 -7.444\\
170.0 -6.6129999999999995\\
175.0 -5.0889999999999995\\};
\addlegendentry{E-plane}
\addplot [semithick, color1]
table [row sep=\\] {%
-180.0 -4.197\\
-175.0 -4.284\\
-170.0 -4.538\\
-165.0 -4.9639999999999995\\
-160.0 -5.5729999999999995\\
-155.0 -6.382999999999999\\
-150.0 -7.421\\
-145.0 -8.732999999999999\\
-140.0 -10.39\\
-135.0 -12.52\\
-130.0 -15.31\\
-125.00000000000001 -18.96\\
-119.99999999999999 -21.97\\
-115.0 -19.74\\
-110.0 -15.91\\
-105.00000000000001 -12.87\\
-100.0 -10.49\\
-95.0 -8.549\\
-90.0 -6.9\\
-85.0 -5.452000000000001\\
-80.0 -4.15\\
-75.0 -2.96\\
-70.0 -1.86\\
-65.0 -0.838\\
-59.99999999999999 0.1131\\
-55.0 0.9962\\
-50.0 1.8119999999999998\\
-45.0 2.56\\
-40.0 3.238\\
-35.0 3.844\\
-29.999999999999996 4.375\\
-25.0 4.831\\
-20.0 5.207000000000001\\
-14.999999999999998 5.502999999999999\\
-10.0 5.716\\
-5.0 5.845\\
0.0 5.888999999999999\\
5.0 5.848\\
10.0 5.722\\
14.999999999999998 5.511\\
20.0 5.216\\
25.0 4.84\\
29.999999999999996 4.3839999999999995\\
35.0 3.8510000000000004\\
40.0 3.2430000000000003\\
45.0 2.5610000000000004\\
50.0 1.81\\
55.0 0.9898\\
59.99999999999999 0.102\\
65.0 -0.854\\
70.0 -1.881\\
75.0 -2.987\\
80.0 -4.1819999999999995\\
85.0 -5.49\\
90.0 -6.944\\
95.0 -8.597999999999999\\
100.0 -10.54\\
105.00000000000001 -12.92\\
110.0 -15.95\\
115.0 -19.69\\
119.99999999999999 -21.73\\
125.00000000000001 -18.78\\
130.0 -15.21\\
135.0 -12.45\\
140.0 -10.34\\
145.0 -8.693\\
150.0 -7.385\\
155.0 -6.35\\
160.0 -5.545\\
165.0 -4.941\\
170.0 -4.522\\
175.0 -4.276\\};
\addlegendentry{H-plane}
\end{polaraxis}

\end{tikzpicture}
    }
    \caption{Simulated H- and E-plane of the patch antenna in \si{dBi}.}%
    \label{fig:radiation-pattern}
\end{figure}

\begin{figure}[tbp]
    \centering
    \includegraphics[width=.9\linewidth]{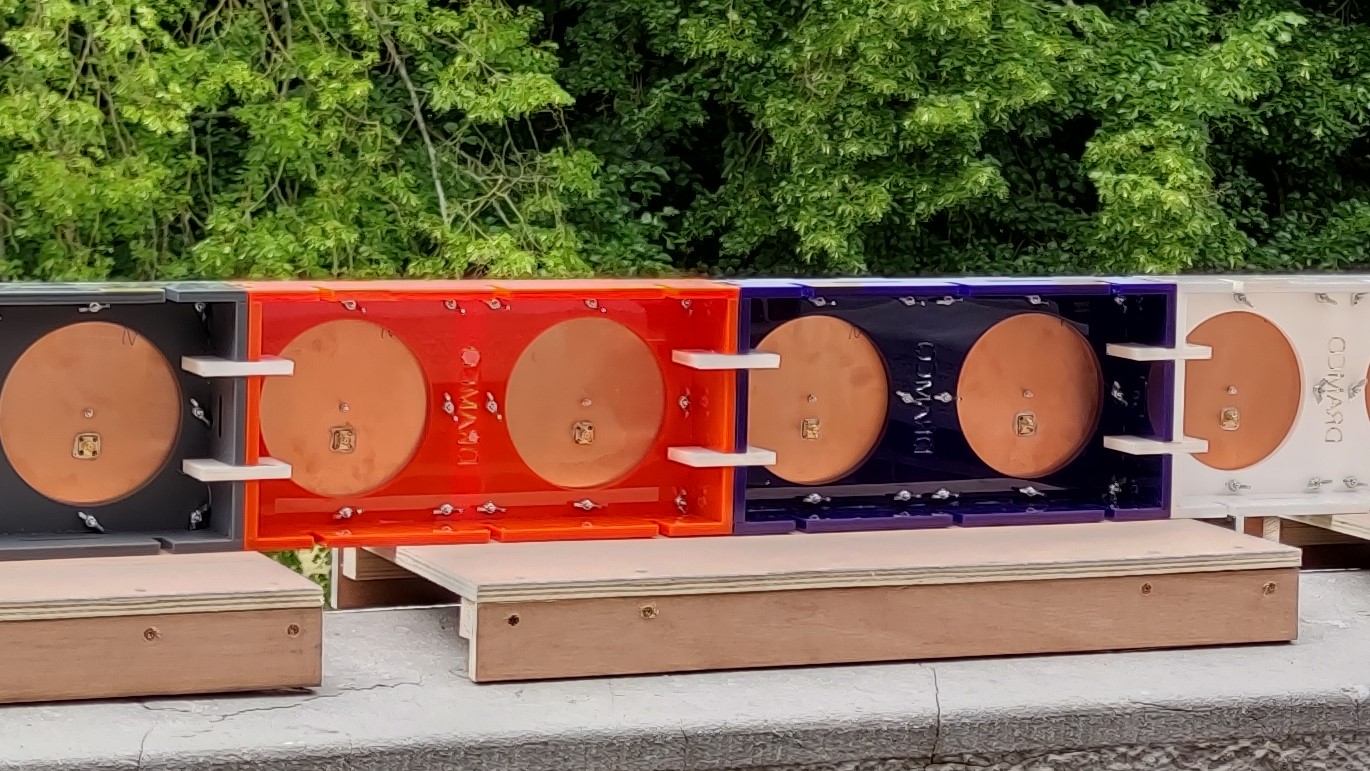}
    \caption{Multiple modular antenna holders connected.}%
    \label{fig:antennas-back}
\end{figure}

\begin{figure}[tbp]
     \centering
     \begin{subfigure}{0.9\linewidth}
         \centering
         \includegraphics[width=\textwidth]{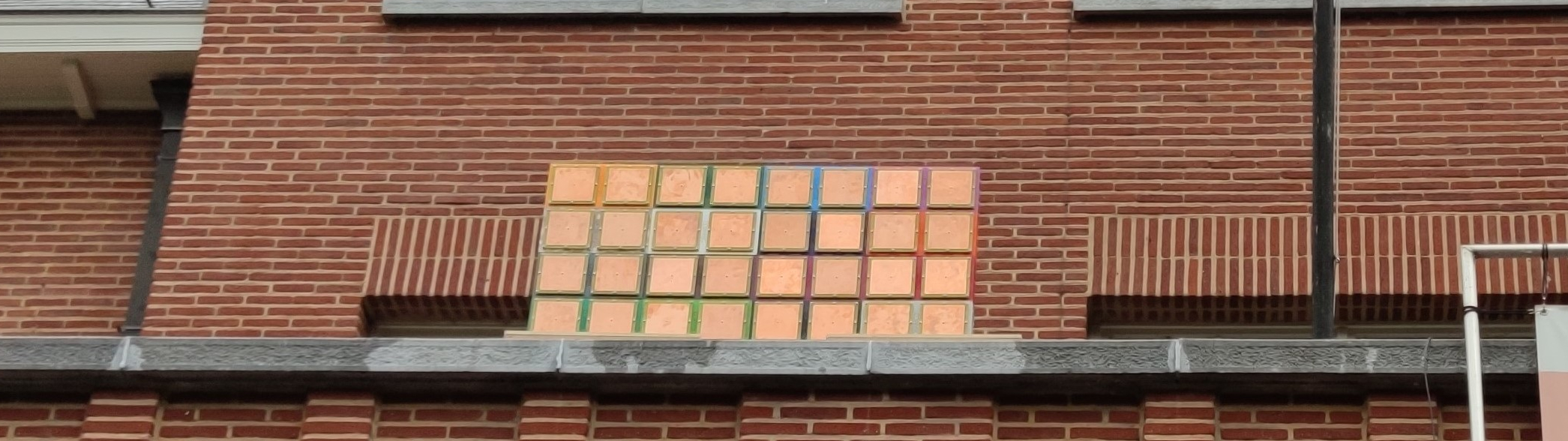}
         \caption{4 \(\times\) 8 URA}\label{fig:ura}
     \end{subfigure}
     
    \begin{subfigure}{0.9\linewidth}
         \centering
         \includegraphics[width=\textwidth]{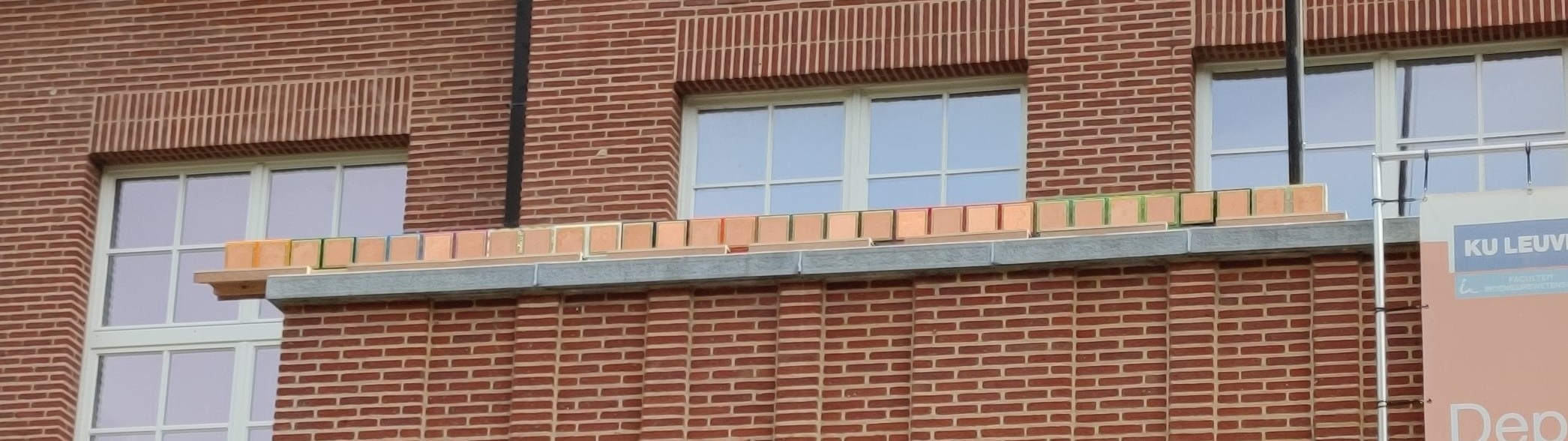}
         \caption{32-element ULA}\label{fig:ula}
     \end{subfigure}
     \caption{Antenna array configurations.}\label{fig:conf}
\end{figure} 

\section{Versatile Large Array}\label{sec:array}

A flexible patch antenna array was developed in-house and deployed at the base station side. The patch was designed for the unlicensed frequency band \SIrange{865.00}{869.65}{\mega\hertz}. In this band, an \(S_{1,1}\) lower than \SI{-10}{\deci\bel} was targeted. As can been seen from Figure~\ref{fig:s11}, the \(S_{1,1}\) is lower than \SI{-10}{\deci\bel} between \SIrange{856.25}{874.25}{\mega\hertz} and is hence meeting the specifications. To lower the impact of mutual coupling between these antennas~\cite{8046083}, patch antennas were preferred over dipoles. The radiation pattern of the patch antenna is shown in Fig.~\ref{fig:radiation-pattern}. The antenna has a gain of \SI{5.9}{\deci\bel i}. 
Two patch antennas are contained in one antenna holder. The array is made up of antenna holders, where each holder has two patch antennas. The holders were designed to be modular such that different configurations could be easily assembled\footnote{\url{dramco.be/massive-mimo/sub-ghz-array/animation.gif}}; the back side of these holders and how to connect them can be seen in Fig.~\ref{fig:antennas-back}. Furthermore, our design enforces that the antennas are always spaced by half a wavelength, independent of the the manner in which the holders are connected, i.e., vertically or horizontally.
For the experiments we targeted a \num{32} antenna ULA but also a URA with the \num{32} antennas split in four rows of eight
antennas each. These configurations are shown in Fig.~\ref{fig:conf}. Due to the modular design of the antenna holders, many other configurations could be made including, but not limited to, cylindrical and distributed setups.

\section{Transmission in the Unlicensed Sub-GHz Band}\label{sec:regulations}
In order to comply to the regulations of short-range devices (SRD860), the utilized frequency band, the transmit power, bandwidth and duty cycle of the existing massive MIMO framework needed to be adapted. An overview of these limitations are given in~\cite{8630442}. These restrictions apply to specific frequency bands and applications. The allowed bandwidth in the SRD band varies between \SI{5}{\kilo\hertz} and \SI{5}{\mega\hertz}. In these bands a maximum transmit power and duty cycle are also specified. The duty cycle ranges from 0.1\% (\SI{3.6}{\second} per hour) to 10\% (\SI{36}{\second} per hour). The maximum transmit power is defined as the Effective Radiated Power (ERP). The ERP is the total power that would have been fed to a half-wave dipole to get the same radiation intensity as the actual device at the same distance and in the direction of the antenna's strongest beam. Hence, the ERP can be expressed as
\begin{equation}
    \mathrm {ERP} =G_{\text{d}}\, P_{\text{in}} \enspace ,
\end{equation}
where \(G_{\text{d}}\) is the gain of the actual antenna compated to a reference half-wave antenna and \(P_{\text{in}}\) the input power.
The ERP is limited between \SI{5}{\milli\watt} (\SI{7}{\dBm}) and \SI{2000}{\milli\watt} (\SI{33}{\dBm}). 

In this work, we use the SRD band 54\footnote{European Union, Commission Decision of 9 November 2006 on harmonisation of the radio spectrum for use by short-range devices (2006). 2006/771/EC. Consolidated version of August 2017.}. The band is defined between \SI{869.40}{\mega\hertz} and \SI{869.65}{\mega\hertz}, and limits the ERP to \SI{500}{\milli\watt} (\SI{27}{\dBm}) and the duty cycle to 10\%.  We use a carrier frequency of \SI{869.525}{\mega\hertz}. 

\textbf{Transmit Power.} As the base station in this measurement setup is only acting as a receiver, the only transmit power that needs to be restricted is the one of the node. This is easily done in the interface of the original framework. The utilized USRP (NI USRP 2952) has a maximum transmit power of \SI{20}{\dBm}, thereby respecting the maximum allowable transmit power of \SI{27}{\dBm} as per SRD860.

\textbf{Bandwidth.} The bandwidth used for real-time operation of the testbed is \SI{20}{\mega\hertz}. Orthogonal frequency-division multiplexing (OFDM) is used and as in LTE, \num{1200} subcarriers are used for carrying the data, which is further divided into \num{100} resource blocks and a subcarrier spacing of \SI{15}{\kilo\hertz}. To respect the regulations of having a bandwidth of maximum \SI{250}{\kilo\hertz}, this has to be reduced. Implementation-wise this was solved by transmitting zeros on all subcarriers except for the \num{13} 
subcarriers in the middle of the symbols, resulting in a total bandwidth of \SI{195}{\kilo\hertz}. This means that uplink pilots were transmitted at two subcarriers in an OFDM symbol and that one resource block of data was utilized.

\begin{figure}[tbp]
    \centering
    \includegraphics[width=0.9\linewidth]{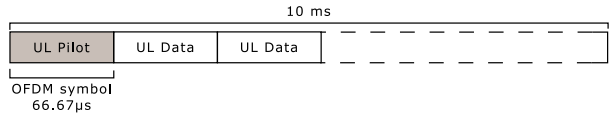}
    \caption{Modified time frame with three OFDM symbols per \SI{10}{\milli\second} slot.}%
    \label{fig:time-sync}
\end{figure}

\textbf{Duty Cycle.}
The maximum duty cycle, i.e., transmit-to-silent ratio, depends on the radio frequency (RF) band. The duty cycle is computed as 
\begin{equation}
    \text{DC}_{max} = \frac{\sum T_{on}}{T_{obs}} \enspace ,
\end{equation}
over a time window of one 
hour. In order to respect the duty cycle regulations, the frame structure had to be changed as well. The default frame structure is based on LTE with \num{10} subframes, each with two slots where each slot is containing
seven OFDM symbols. For all experiments, the first OFDM symbol in the first slot was an uplink pilot.
The two symbol following symbols were set to be transmitting uplink data, as in Fig.~\ref{fig:time-sync} where the first three symbols of the the first slot in the frame structure is shown. All other symbols were empty yielding a duty cycle of 2\% (\(3 \times \SI{66.67}{\micro\second} / \SI{10}{\milli\second}\)).

The framework featured Over-the-Air (OTA) synchronization.
Due to the modification the original slots dedicated for synchronization were removed. Atomic clocks were instead
used for syncing the base station and the mobile station. 
Therefore, the synchronization source in the framework
needed to be changed from the local oscillator to the input clock reference where the atomic clock is connected.

\section{Measurement campaign and Open-Source Data}\label{sec:scenario}
\begin{figure}[tbp]\centering
\fontsize{6pt}{10pt}\selectfont
    \def\svgwidth{0.9\columnwidth}
\begingroup%
  \makeatletter%
  \providecommand\color[2][]{%
    \errmessage{(Inkscape) Color is used for the text in Inkscape, but the package 'color.sty' is not loaded}%
    \renewcommand\color[2][]{}%
  }%
  \providecommand\transparent[1]{%
    \errmessage{(Inkscape) Transparency is used (non-zero) for the text in Inkscape, but the package 'transparent.sty' is not loaded}%
    \renewcommand\transparent[1]{}%
  }%
  \providecommand\rotatebox[2]{#2}%
  \newcommand*\fsize{\dimexpr\f@size pt\relax}%
  \newcommand*\lineheight[1]{\fontsize{\fsize}{#1\fsize}\selectfont}%
  \ifx\svgwidth\undefined%
    \setlength{\unitlength}{971.9502088bp}%
    \ifx\svgscale\undefined%
      \relax%
    \else%
      \setlength{\unitlength}{\unitlength * \real{\svgscale}}%
    \fi%
  \else%
    \setlength{\unitlength}{\svgwidth}%
  \fi%
  \global\let\svgwidth\undefined%
  \global\let\svgscale\undefined%
  \makeatother%
  \begin{picture}(1,0.83844001)%
    \lineheight{1}%
    \setlength\tabcolsep{0pt}%
    \put(0,0){\includegraphics[width=\unitlength,page=1]{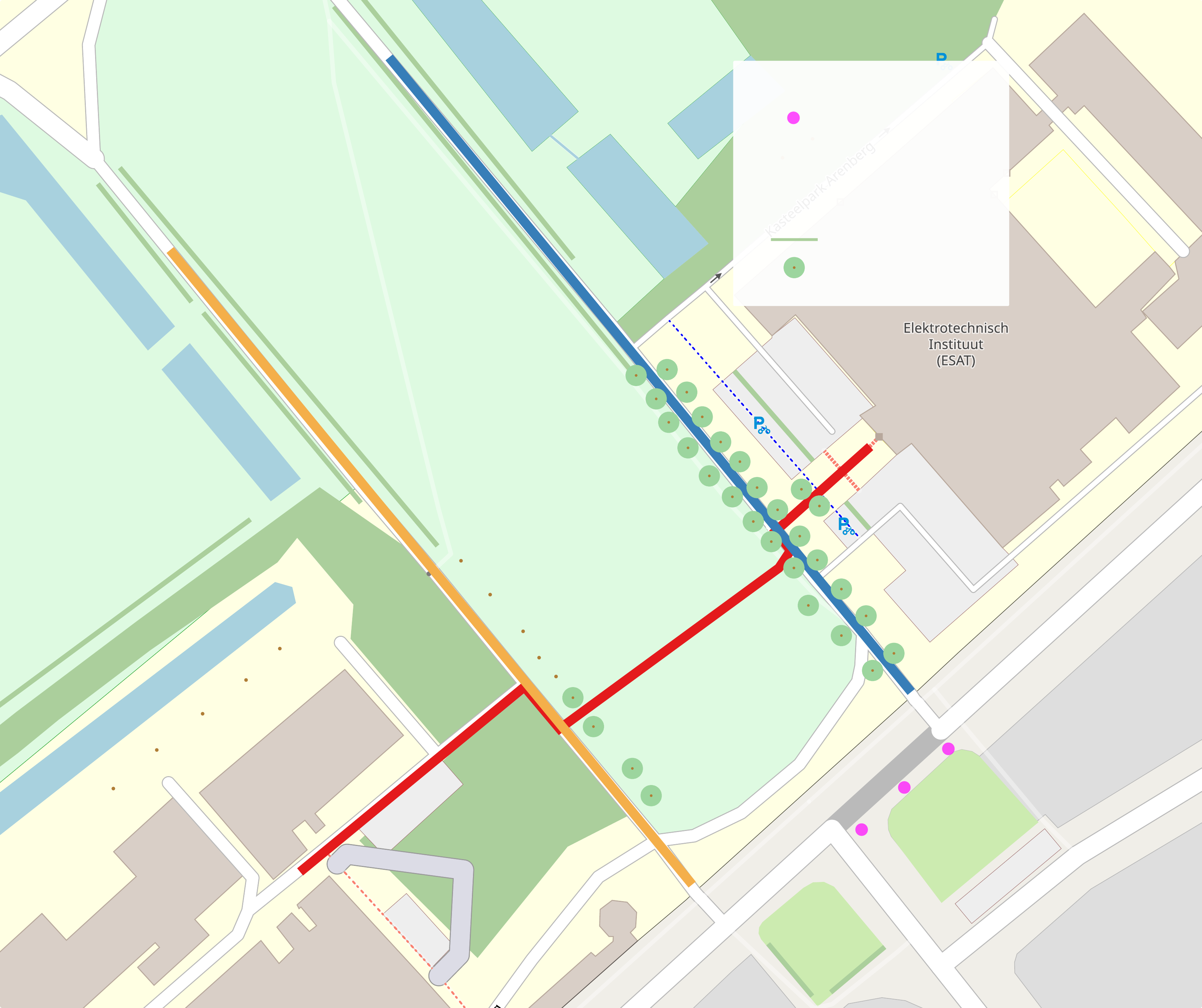}}%
    \put(0.69961805,0.7590149){\makebox(0,0)[lt]{\lineheight{1.25}\smash{\begin{tabular}[t]{l}Base Station\end{tabular}}}}%
    \put(0.69961805,0.73451164){\makebox(0,0)[lt]{\lineheight{1.25}\smash{\begin{tabular}[t]{l}Points D\end{tabular}}}}%
    \put(0.69961805,0.71000841){\makebox(0,0)[lt]{\lineheight{1.25}\smash{\begin{tabular}[t]{l}Path A\end{tabular}}}}%
    \put(0.69961805,0.68550512){\makebox(0,0)[lt]{\lineheight{1.25}\smash{\begin{tabular}[t]{l}Path B\end{tabular}}}}%
    \put(0.69961805,0.66100189){\makebox(0,0)[lt]{\lineheight{1.25}\smash{\begin{tabular}[t]{l}Path C\end{tabular}}}}%
    \put(0,0){\includegraphics[width=\unitlength,page=2]{map_v2.pdf}}%
    \put(0.69851352,0.63524049){\makebox(0,0)[lt]{\lineheight{1.25}\smash{\begin{tabular}[t]{l}Hedge\end{tabular}}}}%
    \put(0.69792825,0.61000076){\makebox(0,0)[lt]{\lineheight{1.25}\smash{\begin{tabular}[t]{l}Tree\end{tabular}}}}%
    \put(0,0){\includegraphics[width=\unitlength,page=3]{map_v2.pdf}}%
  \end{picture}%
\endgroup%
\caption{Overview of rural measurement area. All paths have each a total length of approximately \SI{140}{\meter}.}\label{fig:map}
\end{figure}

\begin{figure}[tbp]
    \centering
    \includegraphics[width=0.9\linewidth]{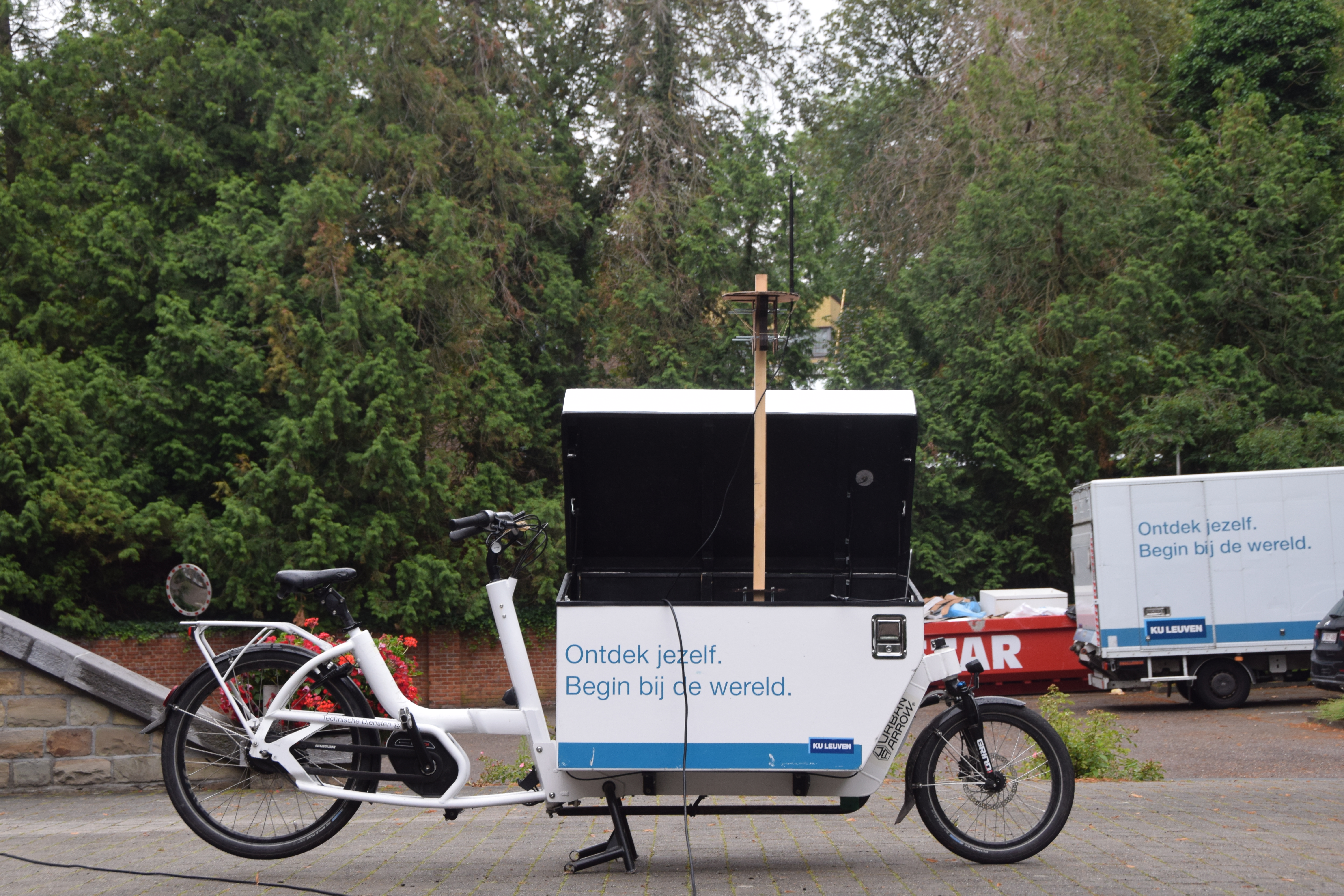}
    \caption{Transmit node.}%
    \label{fig:ue}
\end{figure}

The measurement campaign was performed in front of the Department of Electrical Engineering (ESAT) in Heverlee, Belgium. An overview of this environment can be seen in Fig.~\ref{fig:map}. The base station was equipped with either an ULA or an URA, as shown in Fig.~\ref{fig:ula} and \ref{fig:ura}, and placed on the balcony of the building at a height of 7~meters. Meanwhile the node, seen in Fig.~\ref{fig:ue}, was moving around along the paths that are also depicted on the map. Measurements were collected for all 32 antennas\footnote{Due to an unexpected issue, only the first 31 channels could be used.} at static points, spaced 10 meters apart, and continuously along the same paths. The chosen paths include positions both in Line-of-Sight (LoS) and non-LoS (NLoS) and also paths that are either perpendicular or parallel to the base station antenna array. The details of the measurement setup can be found in Table~\ref{tab:measurement-setup} and the measurements are available open-source\textsuperscript{\ref{data}}. 

\begin{table}[tbp]
\centering
\caption{Measurement setup}
\label{tab:measurement-setup}
\centering
  \begin{threeparttable}
\begin{tabular}{@{}lll@{}}
\toprule
Parameter &  Symbol & Value\\ \midrule
Carrier frequency & \(f_c\) & \SI{869.525}{\mega\hertz} \\
Number of subcarriers (\SI{15}{\kilo\hertz}) & \(F\) & 2 \\
Number of snapshots & \(N\) & 1000\tnotex{tnote:static} /6000\tnotex{tnote:continuous} \\
Transmit power (coerced) &\(P_{tx}\)& \SI{22.6}{\dBm}\\
Number of base station antennas &\(M\)& 32 \\
Number of nodes &\(K\)& 1 \\
Base station Array configuration && ULA/URA \\
Type of BS antenna && Patch \\
Type of UE antenna && Dipole \\
Sample interval && \SI{10}{\milli\second}\\
Sample duration && \SI{66.67}{\micro\second}\\
Measurement duration && 10s/60s\\
Subcarrier modulation && QPSK \\ 
BS height && \SI{7}{\meter} \\
UE height && \SI{1.5}{\meter} \\
Antenna polarization && vertical \\\bottomrule
\end{tabular}
\begin{tablenotes}
      \item\label{tnote:static} For static measurements.
      \item\label{tnote:continuous} For continuous measurements.
    \end{tablenotes}
\end{threeparttable}
\end{table}


\section{Experimental Exploration}\label{sec:results}

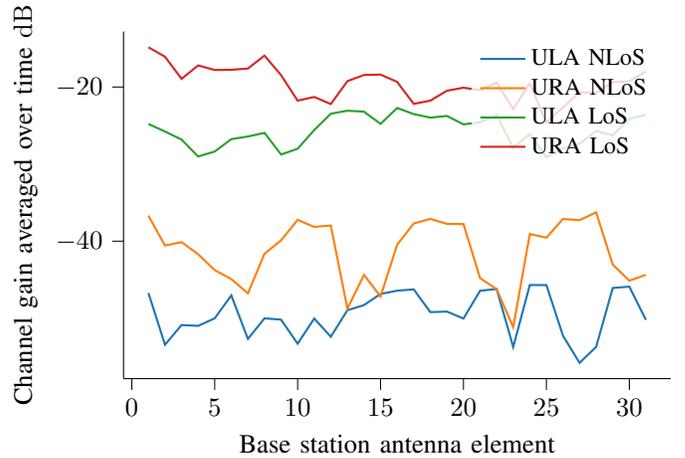
\begin{figure}[tbp]
    \centering
\begin{tikzpicture}

\definecolor{color0}{rgb}{0.12157,0.46667,0.70588}
\definecolor{color1}{rgb}{1.00000,0.49804,0.05490}
\definecolor{color2}{rgb}{0.17255,0.62745,0.17255}
\definecolor{color3}{rgb}{0.83922,0.15294,0.15686}

\begin{axis}[
axis lines* = {left},
tick align=outside,
tick pos=left,
width = \linewidth,
height = .7\linewidth,
x grid style={white!69.01961!black},
xmin=-0.50000, xmax=32.50000,
xtick style={color=black},
y grid style={white!69.01961!black},
ymin=-57.83839, ymax=-12.80121,
ytick style={color=black},
legend style={fill opacity=0.8, draw opacity=1, text opacity=1, draw=white!80.00000!black},
legend style={nodes={scale=0.9, transform shape}},
legend style={draw=none},
legend cell align={left},
xlabel={Base station antenna element},
ylabel={Channel gain averaged over time \si{\deci\bel}}
]
\addplot [color0]
table [row sep=\\] {%
1.00000 -46.71358\\2.00000 -53.43798\\3.00000 -50.88284\\4.00000 -50.98810\\5.00000 -50.00760\\6.00000 -47.03795\\7.00000 -52.68644\\8.00000 -50.01728\\9.00000 -50.18218\\10.00000 -53.30617\\11.00000 -50.03649\\12.00000 -52.40016\\13.00000 -48.96370\\14.00000 -48.29929\\15.00000 -46.84654\\16.00000 -46.42740\\17.00000 -46.25505\\18.00000 -49.20893\\19.00000 -49.13705\\20.00000 -50.04067\\21.00000 -46.43873\\22.00000 -46.20297\\23.00000 -53.72695\\24.00000 -45.69391\\25.00000 -45.69677\\26.00000 -52.26676\\27.00000 -55.79124\\28.00000 -53.71633\\29.00000 -46.07596\\30.00000 -45.89877\\31.00000 -50.21259\\};
\addlegendentry{ULA NLoS}
\addplot [color1]
table [row sep=\\] {%
1.00000 -36.68320\\2.00000 -40.56805\\3.00000 -40.12603\\4.00000 -41.70012\\5.00000 -43.75961\\6.00000 -44.92354\\7.00000 -46.77251\\8.00000 -41.64088\\9.00000 -39.89583\\10.00000 -37.23055\\11.00000 -38.15516\\12.00000 -37.96914\\13.00000 -48.75961\\14.00000 -44.38003\\15.00000 -47.17915\\16.00000 -40.45967\\17.00000 -37.70943\\18.00000 -37.11354\\19.00000 -37.75654\\20.00000 -37.78000\\21.00000 -44.79256\\22.00000 -46.23595\\23.00000 -51.16109\\24.00000 -39.07554\\25.00000 -39.54397\\26.00000 -37.12661\\27.00000 -37.27601\\28.00000 -36.26528\\29.00000 -43.03911\\30.00000 -45.12882\\31.00000 -44.36856\\};
\addlegendentry{URA NLoS}
\addplot [color2]
table [row sep=\\] {%
1.00000 -24.78936\\2.00000 -25.77691\\3.00000 -26.81540\\4.00000 -29.01305\\5.00000 -28.37051\\6.00000 -26.75989\\7.00000 -26.42576\\8.00000 -25.94628\\9.00000 -28.75709\\10.00000 -28.00812\\11.00000 -25.58139\\12.00000 -23.46432\\13.00000 -23.07509\\14.00000 -23.20042\\15.00000 -24.78502\\16.00000 -22.72066\\17.00000 -23.51471\\18.00000 -23.97171\\19.00000 -23.75473\\20.00000 -24.85023\\21.00000 -24.57717\\22.00000 -23.62233\\23.00000 -27.90920\\24.00000 -26.06665\\25.00000 -29.13476\\26.00000 -27.99005\\27.00000 -27.35372\\28.00000 -25.70577\\29.00000 -26.28103\\30.00000 -24.14348\\31.00000 -23.59443\\};
\addlegendentry{ULA LoS}
\addplot [color3]
table [row sep=\\] {%
1.00000 -14.84835\\2.00000 -16.07139\\3.00000 -18.94107\\4.00000 -17.20110\\5.00000 -17.79635\\6.00000 -17.76586\\7.00000 -17.60549\\8.00000 -15.92417\\9.00000 -18.46371\\10.00000 -21.77430\\11.00000 -21.29078\\12.00000 -22.20798\\13.00000 -19.24189\\14.00000 -18.43727\\15.00000 -18.38891\\16.00000 -19.34116\\17.00000 -22.18844\\18.00000 -21.76821\\19.00000 -20.48675\\20.00000 -20.08749\\21.00000 -20.40656\\22.00000 -19.42819\\23.00000 -22.85117\\24.00000 -19.53380\\25.00000 -24.62375\\26.00000 -22.62800\\27.00000 -20.69124\\28.00000 -20.72975\\29.00000 -19.37398\\30.00000 -19.26287\\31.00000 -17.98381\\};
\addlegendentry{URA LoS}
\end{axis}

\end{tikzpicture}
    \caption{The average channel gain per antenna for the two array configurations in LoS and NLoS. The antenna elements are numbered from right to left and from the bottom to the top.}
    \label{fig:avg-gain}
\end{figure}

In this section a first assessment of the channel conditions of measured massive MIMO at sub-GHz frequencies is presented. The antenna elements are numbered from right to left and from the bottom to the top. A first look of what the average received channel gain for the different antennas can look like for an ULA and an URA in both LoS and NLoS is shown in Fig.~\ref{fig:avg-gain}. The channel gain per antenna at position \(k\) is averaged over time and frequency as in
\begin{equation}\label{eq:combining-gain}
\frac{1}{N F} 
\sum^{N, F}{
\abs{\vectr{h}_{k,m}(n,f)}^2} \enspace ,
\end{equation}
where \(\vectr{h}_{k,m}(n,f)\) is the channel vector measured at position k at snapshot \(n\), frequency \(f\) and at antenna \(m\). By averaging the channel gain over time and frequency, the small-scale effect is averaged out and thus Fig.~\ref{fig:avg-gain} depicts the large-scale fading coefficient per antenna. The differences between these large-scale fading coefficients per antenna illustrate the large-scale fading over the array.
The LoS and NLoS points are chosen as one static point in the beginning of path~A, close to the base station array, and one static point 100~meters away along the same path, respectively. Due to the different path losses, the NLoS curves will naturally be lower than the LoS ones. Another difference that can be observed is that for the LoS cases the average channel gain is more similar for the different antennas in both arrays while in NLoS a more prominent large-scale fading effect can be observed over the array; this is especially clear for the URA where four dips of channel gain can be observed -- one for each row in the array. Furthermore, the variation of channel gain per antenna is also larger for the NLoS points. 
The difference between the standard deviation of channell gain per antenna for the ULA between the LoS and NLoS scenario is \SI{1}{\deci\bel} and the corresponding value for the URA is \SI{1.8}{\deci\bel}. The URA experiences a \SI{1.1}{\deci\bel} higher standard deviation than the ULA in the NLoS scenario. Finally, one last observation from Fig.~\ref{fig:avg-gain} is that the URA in general has a higher average channel gain than the ULA.

\begin{figure}[tbp]
    \centering
    \input{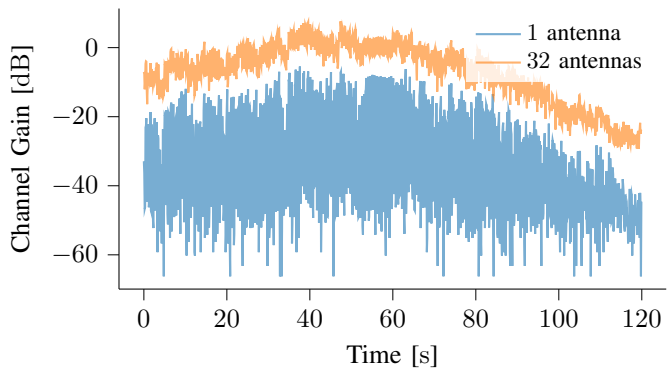}
    \caption{Channel gain over time (along path B) for \num{1} and \num{32} base station antennas in the ULA, respectively.}
    \label{fig:chhard-ULA-B}
\end{figure}

One advantage with massive MIMO is the array gain resulting from combining the many antennas, resulting in a potential transmit power reduction at the node side. Moreover, as the number of base station antennas $M$ increases, the variation of channel gain decreases; this phenomena is called channel hardening and makes a fading channel behave more deterministic. We adopt the definition given in~\cite{downlink_pilots}, stating that a channel $\mathbf{h}_{k}$ offers channel hardening when the number of base station antennas $M$ goes to infinity if

\begin{equation}
\frac{\var{\|\mathbf{h}_{k}\|^2}}{\expt{\|\mathbf{h}_{k}\|^2}^2}\rightarrow 0, \hspace{0.5cm} \text{as} \hspace{0.2cm} M\rightarrow \infty,
\label{eq:chhard_def}
\end{equation}

\noindent where $\var{.}$ is the variance, $\expt{.}$ the expectation and $\|.\|$ is the Euclidean norm. Both the array gain and the channel hardening can be observed in the experimental results depicted in Fig.~\ref{fig:chhard-ULA-B} where the lower curve corresponds to one antenna and the upper curve is when combining all antennas for the ULA along the continuous path~B measurement. Due to the array gain, the latter is moved up and the channel hardening effect can be seen as the variations of channel gain becoming insignificant in comparison to the one antenna case. Both the increased received gain and the smaller fading margin required can reduce the transmit power needed at the node side.

\section{Conclusions}\label{sec:conclusions}
In this work, we have presented an experimental setup and the first measurement campaign using massive MIMO at sub-GHz frequencies and made the data available open-source. Our initial assessment shows the potential benefits of using massive MIMO for connecting remote energy-constrained devices. 
Due to the presence of channel hardening, the reliability of the link can be increased. In addition, the array gain allows to decrease the transmit power of the devices, or to extend the coverage. The impact of using an ULA in comparison to an URA in LoS and NLoS conditions is investigated. Our study demonstrates the feasibility of using massive MIMO at sub-GHz frequencies to support future LPWANs. These measurements enable further investigation of, e.g., precoding and scheduling algorithms for IoT devices. 




\section{Acknowledgment}
We would like to thank Vladimir Volski for designing the patch antenna and our colleagues in Dramco for assisting us during the experiments. 

 \FloatBarrier
\bibliographystyle{IEEEtranN}
{
\bibliography{bib}}

\end{document}